# Numerical analysis of partial discharge ignition in H$_2$ bubbles floating in dielectric oils, for High-Voltage Solid State Transformer applications

Konstantinos Kourtzanidis, Panagiotis Dimitrakellis, Dimitrios Rakopoulos

*Abstract*—We report on a self-consistent numerical analysis campaign of partial discharge (PD) ignition in H$_2$ bubbles floating in biobased dielectric oils. We investigate various configurations (bubble sizes, bubble position, existence of protrusion) on a cylinder-to-cylinder setup that emulates a specific SST module (from SSTAR Horizon Europe project) under transient overvoltage as well as in its design operational conditions (V$_{RMS}$ = 66 kV, AC excitation of 50 Hz). Our results on electrical characteristics and plasma dynamics leading to the PD ignition, indicate that under transient overvoltage and for mm size bubbles (diameter 1 - 4.5 mm), the smaller the bubble the less the inception voltage, while the peak inception voltage is higher than 70 kV. The existence of metallic protrusion can affect the inception voltage of a remote floating bubble only slightly and when this is close to the sharp tip. The extreme scenario of a protrusion in contact (inside) a gas bubble severely affects the insulation properties and drops the PD inception voltage remarkably. The larger the bubble and the sharper the tip of the protrusion the lower the inception peak voltage, that can reach values well below 40 kV. On the contrary and under design operation, larger bubbles increase the severity and probability of PD events, leading to lower instantaneous inception voltages. Current pulses produced in bubbles can quickly transit to intense streamer discharges (which can also transit to catastrophic arcing) if the operational frequency is reduced and/or under transient, HF overvoltage.

## I. INTRODUCTION

TYPICALLY, more than 70% of transformer faults are internal, initiated via partial discharges inside the transformer insulation. The initial discharge can cause gradual damage to the insulating system and lead to a complete breakdown [1]. In high-voltage (HV) transformers, oils are typically used as isolation systems. The main cause of transformer failure is the incipient (partial) discharges that occur due to defects in the insulation system and can gradually damage the insulation and lead to breakdown [2]. Defects are any gas cavities (bubbles), protrusions in the winding or floating particles that cause local field enhancement and facilitate partial discharge (PD) ignition. PDs may be of different type, e.g., internal discharges that occur due to the presence of bubbles inside the dielectric medium, surface discharges caused by the lower dielectric strength at the interfaces, corona discharges at sharp metallic points of elevated electric field, and dielectric barrier discharges (DBDs) associated with insulating materials (dielectric fluids) between metallic terminals.

The presence of bubbles in a dielectric fluid increases the breakdown probability and thus several studies have been focused on such phenomena [3], [4]. It is widely accepted that the initial stage of the transformer oil breakdown is caused by a gas bubble formed by evaporation due to local heating in high electric field regions of the electrode surface [5]. In addition, stray gassing (the formation of gases in electrical insulating oils heated at relatively low temperatures) favor the production of saturated carbon gas bubbles and especially hydrogen (H$_2$) in temperatures below 120 °C [6], [7]. H$_2$ production is even more likely to occur in synthetic esters and vegetable oils, which are gaining attention for transformer applications, and where a great deal of water (up to 1500 ppm) can be solubilized, reducing the formation of vapor bubbles.

While several experimental and theoretical studies on bubbles immersed in dielectric liquids exist [5], PD modeling approaches can enable a fast and in-depth understanding of the main design parameters that can lead to insulation failure. As the PD is most probable when small amounts of gases are present, several models [8] have been developed to simulate the PD in voids (e.g., gas bubbles) such as capacitance models [9], electrostatic or induced charge models [10], [11], and conductance models [12]. Plasma models including corona and especially self-consistent plasma-fluid models, the latter simultaneously solving the Poisson equation with the plasma drift-diffusion-reaction system, significantly relax the main and inherent assumptions of these approaches on the PD occurrence inside cavities. However, numerical studies on PD inception in gas bubbles are scarce [13]–[15].

Herein we employ a plasma-fluid model to study PD inception at conditions that cannot be identified at routine tests, considering extreme conditions (e.g., high frequency, overvoltage) and worst-case scenarios in terms of insulation system failure (e.g., defects). We use self-consistent plasma description and explore different scenarios of PD inception considering a static H$_2$ bubble trapped inside transformer oil. We study the effect of additional defects, i.e., a metallic protrusion of high local electric field and we explore both DC and AC high voltage signals.

## II. MODEL DESCRIPTION

We leverage self-consistent plasma models in COMSOL Multiphysics and a house-built (initially developed by ONERA, The French Aerospace Lab) multispecies self-consistent solver (COPAIER). The physical models for both solvers are similar [16],[17]. COPAIER solver incorporates adaptable time-stepping, implicit time-integration [18] for the diffusion-advection-reaction system obeying the dielectric relaxation time scales, efficient finite element (for Poisson equation) and finite volume (for continuity equations) algorithms and MPI parallelism. The Scharfetter-Gummel flux scheme is used here to discretize the convection-diffusion operator of species continuity equations. COPAIER allows the user to choose different schemes and physical models, providing flexibility in the physical, spatial, and temporal order of accuracy and consequently simulation cost. The conservative form of COPAIER solver regarding the drift-diffusion equations provides superior robustness and allows to capture interesting phenomena such as corona/avalanche-to-streamer transitions.

### A. Governing equations

The electron density $n_e$ is computed from the electron density continuity equation with the drift-diffusion approximation:

$$\frac{\partial}{\partial t}(n_e) + \nabla \cdot \mathbf{\Gamma_e} = R_e - (\mathbf{u} \cdot \nabla)n_e$$

where $u$ is the neutral fluid velocity which is assumed as zero in our simulations. $\mathbf{\Gamma_e}$ is the electron flux given by the drift-diffusion approximation and the following equation:

$$\mathbf{\Gamma_e} = -(\mu_e \cdot \mathbf{E})n_e - \nabla(D_e n_e)$$

In the above equations $\mathbf{u_d}$ is the drift velocity for electrons which is given by

$$\mathbf{u_d} = \mu_e \cdot \mathbf{E}$$

where $E$ is the electric field (V/m) and $\mu_e$ is the electron mobility as derived from the equation:

$$\mu_e = \frac{e}{m_e v_m}$$

where $e$ and $m_e$ are the electron charge and mass, respectively, and $v_m$ the momentum transfer collision frequency. $D_e$ is the electron diffusivity as derived from Einstein's equation:

$$D_e = \frac{\mu_e k_B T_e}{e}$$

According to the local-field approximation (LFE) adopted herein, the electron energy $T_e$ is considered a function of the reduced electric field (E/N). The mean electron energy as well as the electron mobility and diffusivity as functions of the E/N are computed via Bolsig+ solver [19] and provided as inputs (lookup tables) to the plasma model (see also below).

$R_e$ is the total rate of the reactions that contribute to the production and loss of electrons as expressed by:

$$R_e = \sum_{j=1}^{M} x_j k_j N_n n_e$$

where $x_j$ is the molar fraction of the target species for the reaction j, $k_j$ the rate coefficient for reaction j, and $N_n$ the neutral number density. Rate coefficients are also calculated via Bolsig+ for all the important reactions in the plasma and are provided to the model as a function of the E/N.

The following equation is solved for the mass fraction of each one of the non-electron species $k$:

$$\rho \frac{\partial}{\partial t}(w_k) + \rho(\mathbf{u} \cdot \nabla)w_k = \nabla \cdot \mathbf{j_k} + R_k$$

where $\rho$ is the mixture density, $w_k$ the mass fraction of species k, $\mathbf{u}$ the vector of the mass averaged fluid velocity, $R_k$ the reaction rate expression for production and loss of species k and $j_k$ the diffusive flux vector given by the equation:

$$\mathbf{j_k} = \rho w_k \mathbf{V_k}$$

where and $\mathbf{V_k}$ is the multicomponent diffusion velocity for species k that depends on the selected option for the Diffusion Model property. In COMSOL's model we have selected 'Mixture Averaged' model and the $V_k$ is given by the relation:

$$V_k = D_{k,m}\frac{\nabla w_k}{w_k} + D_{k,m}\frac{\nabla M_n}{M_n} + D_k^T \frac{\nabla T}{T} - z_k \mu_{k,m} \mathbf{E}$$

with $D_{k,m}$ the mixture averaged diffusion coefficient defined as:

$$D_{k,m} = \frac{1 - w_k}{\sum_{j \neq k}^{Q} x_j / D_{kj}}$$

and $M_n$ the mean molar mass of the mixture computed as:

$$\frac{1}{M_n} = \sum_{k=1}^{Q} \frac{w_k}{M_k}$$

$T$ is the gas temperature, $D_k^T$ the thermal diffusion coefficient for species $k$, $z_k$ the charge number for species $k$, $E$ is the electric field, and $\mu_{k,m}$ the mixture averaged mobility for species $k$ derived from Einstein's relation.

COPAIER instead solves a similar equation to the drift-diffusion equation provided for electrons, for every ion species. Diffusion coefficients are calculated via Einstein's relation.

The electrostatic field is calculated by Poisson's equation:

$$-\nabla \cdot \varepsilon_0 \varepsilon_r \nabla V = \rho$$

that derives from combining the Gauss's law:

$$\nabla \cdot (\mathbf{\varepsilon E}) = \rho$$

with the definition of potential V:

$$\mathbf{E} = -\nabla V$$

where $E$ the electric field vector, $\varepsilon$ the dielectric permittivity of the medium ($\varepsilon = \varepsilon_0 \varepsilon_r$), and $\rho$ the space charge density computed via the following equation based on the chemistry considered:

$$\rho = q\left(\sum_{k=1}^{N} Z_k n_k - n_e\right)$$

Surface charge accumulation on the oil-gas interface is considered by solving an ODE on the boundary:

$$\frac{\partial \rho_s}{\partial t} = \mathbf{n} \cdot \mathbf{J_i} + \mathbf{n} \cdot \mathbf{J_e}$$

where $\rho_s$ is the surface charge density, $n$ is the surface normal vector, $J_i$ and $J_e$ are the total ion current density and total electron current density respectively.

### B. Boundary conditions

We assume that all positive ions are quenched to neutral species upon impingement with the walls (metallic electrode or oil surface) and no reflection occurs. For all the species we assume a sticking coefficient equal to unity. For electrons, secondary emission flux is considered which is described:

$$\mathbf{n} \cdot \mathbf{\Gamma_e} = \left(\frac{1}{2}v_{e,th}n_e\right) - \sum_p \gamma_p \left(\mathbf{\Gamma_p} \mathbf{n}\right)$$

where $v_{e,th}$ is the thermal velocity defined as:

$$v_{e,th} = \sqrt{\frac{8k_b T_e}{\pi m_e}}$$

and $\gamma_p$ is the secondary emission coefficient which is fixed at 0.01 in metal contacts and 0 in dielectric contact as in the case of oil surface where we assume that no secondary emission occurs on the gas-liquid interface.

TABLE I
$H_2$ CHEMISTRY CONSIDERED IN THE SIMULATION

| Reaction | | Threshold Energy (eV) | Rate coefficient | Ref. |
|---|---|---|---|---|
| R1 | $e^- + H_2 \rightarrow H_2^+ + 2e^-$ | 15.4 | $f(T_e)$ | [20] |
| R2 | $H_2^+ + H_2 \rightarrow H_3^+ + H$ | - | $2.1 \times 10^{-9}$ | [21],[22] |
| R3 | $e^- + H_2^+ \rightarrow 2H$ | - | $5.66 \times 10^{-8} \, T_e^{-0.5}$ | [21],[22] |
| R4 | $e^- + H_3^+ \rightarrow H_2 + H$ | - | $9.75 \times 10^{-8} \, T_e^{-0.5}$ | [21],[22] |

*C. Plasma chemistry*

We adopted a simplified hydrogen kinetics model, presented in TABLE I. It accounts only for the charged species production and consumption, neglecting the kinetics involving H atoms [22]. The excited states of $H_2$ and H are also excluded to further simplify the model. The contribution of H atoms to the formation of $H^+$ ions is negligible [22], while the ionization cross sections for the reaction $e^- + H_2 \rightarrow H^+ + H + 2e$, the main channel for $H^+$ formation, are an order of magnitude lower than the reaction $e^- + H_2 \rightarrow H_2^+ + 2e^-$ (< 1% in the low-energy region) [20]. Based also on the branching ratios (0.93 for the former – 0.07 for the latter), [22] we adopt a total ionization rate coefficient considering only the reaction yielding $H_2^+$ ions (R1), thus neglecting the formation of $H^+$. The model includes $H_3^+$ ions that derive from ion-neutral reaction R2 and are expected to be dominant species in the plasma [21]. $H^-$ ions are formed through the dissociative attachment reaction $e^- + H_2 \rightarrow H + H^-$ that has reaction rate coefficient several orders of magnitude lower than the electron impact ionization, thus their contribution is considered negligible [23], [24].

The electron transport and reaction coefficients are obtained by solving the Boltzmann equation under the two-term approximation in BOLSIG+. Cross sections for $H_2$ used as inputs in BOLSIG+ have been taken from the Morgan database [25], [26] which includes 17 electron-neutral scattering cross sections describing dissociative attachment, effective momentum transfer, rotational, vibrational, and electronic excitation, dissociation, and ionization. We assume a gas composition of 100% $H_2$ at atmospheric pressure and a temperature of 360 K, in the range of transformer operation. The ionization degree is assumed $10^{-6}$, the electron density $10^{19}$ m$^{-3}$ and the gas temperature 360 K. Ion mobilities are specified as 13.7 cm$^2$/V·s for $H_2^+$ and 22.5 cm$^2$/V·s for $H_3^+$ ions as taken from [27]. Diffusion coefficients for ions are calculated by the Einstein relation using the constant mobility values. All neutral species are considered immobile and only the reaction terms are included in the overall neutral species balance. For both COMSOL and COPAIER simulations, a low floor electron and $H_3^+$ density is fixed at $10^9$ m$^{-3}$, considering background ionization from cosmic rays.

*D. Computational aspects and simulation parameters*

The computational domain reflects the transformer part between the windings where the voltage is high and thus PD occurrence is most critical. In the simplified approach the windings are smooth and thus the domain consists of a parallel plate configuration. In the case where we study defects on the winding surface, we consider metallic protrusion. Figure 1 shows the computational approach and simulation domain, considering an $H_2$ bubble trapped in between a high electric field protrusion of the primary winding and the secondary winding and immersed in the transformer oil. A 2-D axisymmetric domain and computational mesh are developed in COMSOL Multiphysics software. For the cases run in our house-built solver (COPAIER), the mesh was created in GMSH software [28]. The primary winding in our axisymmetric geometry is fixed at position z=0, whilst the secondary winding is fixed at z = -50 mm (overall gap 50 mm) and is considered grounded (V0 = 0). A high voltage of variable values is applied on the primary winding (high voltage terminal).

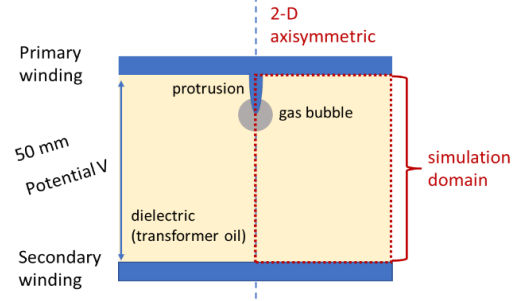

Fig. 1. Computational approach and geometry for a characteristic set of simulations considering a $H_2$ bubble trapped in between the primary and secondary winding of the transformer and in contact with a metallic protrusion with a high electric field.

Considering mm-sized protrusions and typical ellipsoid shapes (with radius of curvature of hundreds of μm as suggested in the literature [30]), we assume elliptical geometry for the protrusion where we fix *a* parameter at 2 mm, while the *b* parameter is a variable [29], [30] (see Fig. 4 and discussion in Sec. III for more details on the geometry aspects). The choice of mm-sized protrusions was made to mimic defects that might arise from machining and polishing of surfaces and to render our numerical approach feasible – smaller defects (in the μm range in length) might be more representative of their actual length. The bubble is assumed spherical (a circle in the 2-D geometry) and the $H_2$ plasma model is applied only inside the bubble. The gas bubble diameter varies between 1 and 4.5 mm; such bubble sizes (corresponding to μL volumes) have been widely reported in experimental studies [31], [32]. The bubble center is positioned at different coordinates in terms of x (radius r) and z, with fixed y = 0 (axis) and a sector angle 180° (half circle). The bubble is placed inside the dielectric fluid (insulation medium) and we consider a fixed relative permittivity value ε = 3, based on literature reports on dielectric properties of most synthetic ester and vegetable oils in the frequency and temperature range considered herein [33]. The temperature is 360 K, in between the optimum transformer operation temperature (70 - 100°C).

We perform two distinct sets of computations regarding the applied voltage waveform on the primary winding: Set A is focusing on high-frequency (HF) simulations in COMSOL and Set B on low-frequency (LF) simulations in COPAIER. For the former, AC applied voltage with a frequency ($f$) of 50 kHz is considered. Set B assumes the LF component to be equivalent to a DC one, ramped up to its maximum value with a long rising time (~ 100 μs) and then kept constant. This is justifiable as plasma dynamics and especially ionization and electron drift timescales are much faster than the LF voltage variations. The simulations of Set B are performed with the in-house solver COPAIER, as COMSOL presented several issues of convergence and numerical errors in these LF regimes while substantial CPU time and memory requirements (owed to the purely finite element formulation of COMSOL equations) was required. COPAIER was more robust, and we leveraged its efficient parallelization, to run the simulations in a multi-processor server machine (>24 nodes available).

*E. Inception voltage*

The currents associated with the PD initiation typically take the form of high-frequency transient pulses. These pulses of ns to μs duration can be repetitive during each half-cycle and/or be reproduced at each subsequent half-cycle of the AC waveform. The PD inception voltage Ui is the lowest voltage at which partial discharges occur in a test circuit when the test voltage is gradually increased. Here the inception voltage is calculated in a twofold way. For Set A, we perform simulations of half an HF-AC cycle for various peak voltages (applied voltage – V0) and track the discharge current versus time. For each current pulse observed, the total charge per pulse is calculated by temporal integration of the discharge current. It is generally assumed that a partial discharge of less than 10 pC doesn't lead to any harm to the insulation. In this numerical campaign though, which deals with the HF 50 kHz source, we choose a stricter threshold of 1 pC. This is due to the expected burst interval between each current pulse (or PD event) which should be in the order of the AC half-period i.e., 1 μs. This burst interval is extremely short and thus accumulation of charges could lead to a critical PD breakdown even for low values of charge per PD event. We thus consider that whenever a current pulse-related total charge exceeds 1 pC, we have PD inception. The inception voltage is then defined as the peak applied voltage of the AC waveform where one current pulse exceeds the charge threshold criterion. For Set B, we also define and identify the instantaneous discharge onset (inception) as the time-instant/instantaneous applied voltage when the slope of the discharge (conduction) current changes (increases) from a smoother "pre-breakdown" like regime to a breakdown / avalanche pulsating (or not) regime and continuous growing. This roughly corresponds to maximum electron densities during the pulse peaks of $10^{13}$ - $10^{14}$ m$^{-3}$. Here we consider that whenever a current pulse-related total charge exceeds 10 pC, we have PD inception. Our definition of the inception voltage (instantaneous or not) is a strict one, as slight deviations from the applied waveform might shift the inception voltage ($V_{inc}$) to slightly larger values and prevent total breakdown and/or partial discharging. The critical PD charge chosen here is also quite strict; a PD of ~10 pC can be rather weak and not detrimental for the SST operation. We consider that the initiation of an even weak PD presents a hazard, as slight overvoltage could easily intensify the discharge and end up into complete breakdown.

### III. RESULTS

*A. Set A: HF simulations in COMSOL*

In Set A we consider a 50 kHz sinusoidal HV signal, and we perform time-dependent calculations for a half period, recording the positive current pulses and induced charges. Based on the considered charge threshold of 1 pC we determine the PD inception voltage for different cases / parametric studies.

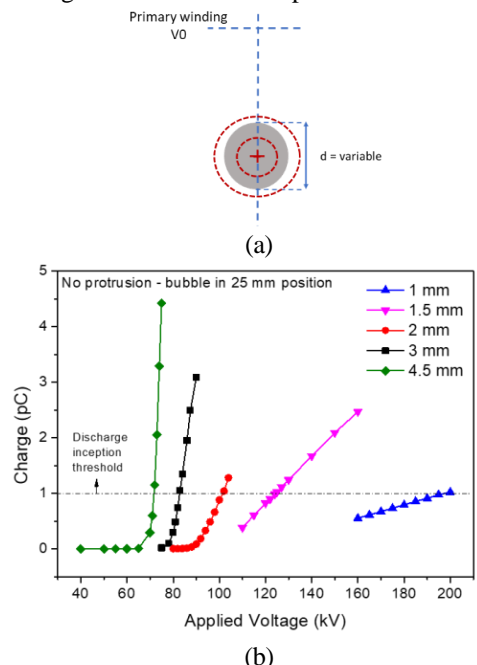

Fig. 2. (a) Computational approach for Set A - Case 1 where a gas bubble of different diameter is placed inside the dielectric fluid at a fixed distance from the primary winding and no protrusion exists. (b) Total charge of the positive pulse as a function of the applied peak voltage $V_0$ for the different bubble sizes (1 – 4.5 mm).

The first parametric study (Case 1) explores the effect of bubble size when the bubble is positioned at the center of the electrode gap inside the dielectric fluid (25 mm). We consider a spherical bubble with its diameter varying from 1 to 4.5 mm. There is no protrusion existing on the winding surface, thus we consider a plane-to-plane geometry that assumes a uniform electric field across the gap. The general approach and geometry considered in this parametric study are depicted in Figure 2a. Based on the considerations about PD inception and the total charge for the case of the HF – 50 kHz signal (Set A), we define the inception voltage for the partial discharge at a strict threshold of 1 pC. We apply different voltage values (peak voltage – V0) and we calculate the induced charge for the positive pulse (half-period). We observe that the 1 pC threshold for PD inception is reached at much lower applied voltage values as the bubble diameter increases (Figure 2b). The increase in the bubble volume leads to a dramatic increase of the inception voltage from around 200 kV peak for the small (d = 1 mm) to less than 80 kV for the large (d = 4.5 mm) bubble. As the accumulation of gas in the form of very small bubbles inside the dielectric fluid will eventually lead to the formation of larger bubbles and dramatically increase the probability of PD occurrence, it is very critical to limit the gas impurities to the minimum possible volume.

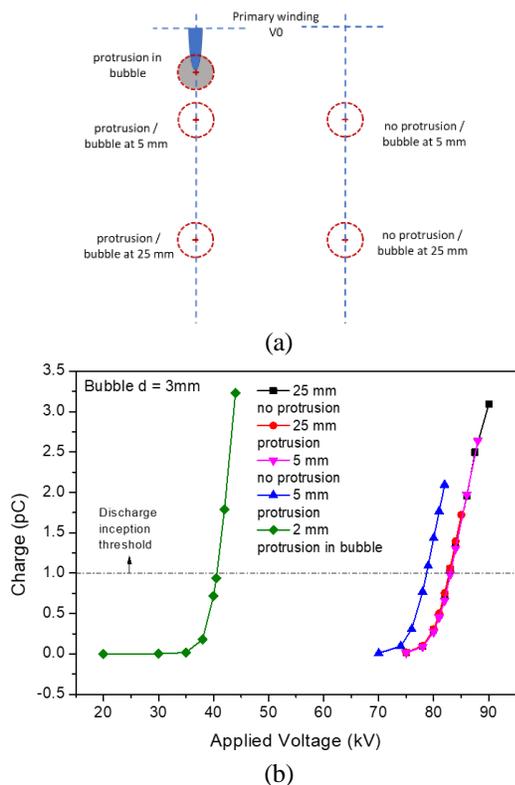

(a)

(b)

Fig. 3. (a) Computational approach for Set A - Case 2 where a gas bubble of fixed diameter 3 mm is placed inside the dielectric fluid at different positions (distance) from the primary winding, with and without protrusion. (b) Total charge of the positive pulse as a function of the applied peak voltage $V_0$ for the different cases.

We also assess the role of bubble position in respect to the primary winding with and without the existence of a sharp protrusion on the primary winding surface (Figure 3a). In this case (Case 2), we aim also to understand the importance of the protrusion existence and its distance from the bubble to the PD inception voltage. We assume elliptical geometry for the protrusion and a spherical geometry for the bubble; the protrusion has a fixed 2 mm length (a-value) and a width at 0.5 mm (b-value). The bubble diameter in this study is 3 mm. Figure 3b shows the variation of the total charge. We clearly see that the relative position of the bubble plays a minor role to the inception voltage as the total charge variation with the voltage is the same either we place the bubble in the middle of the gap (25 mm) or very close to the primary winding of the transformer (5 mm). This applies only to the case where there is no protrusion in the high-voltage electrode, meaning the electric field is uniform across the gap (plane geometry). The inception voltage in both positions is determined at 83 kV. On the contrary, the existence of protrusion does affect the inception voltage, but only if the bubble is quite close to the high-voltage terminal. When the bubble center is positioned at 5 mm distance from the primary winding (3 mm distance from the tip of the protrusion), meaning that the gas-liquid interface is at a low 1.5 mm distance, the PD inception occurs at slightly lower voltage value 79 kV. However, this difference is not very significant, indicating the low probability of inception when there are bubbles floating in the dielectric fluid, considering the high voltage values required. The inception voltage drops dramatically though in the extreme case where the bubble is in contact with the protrusion. We clearly see that the inception voltage drops to around 40 kV in this case.

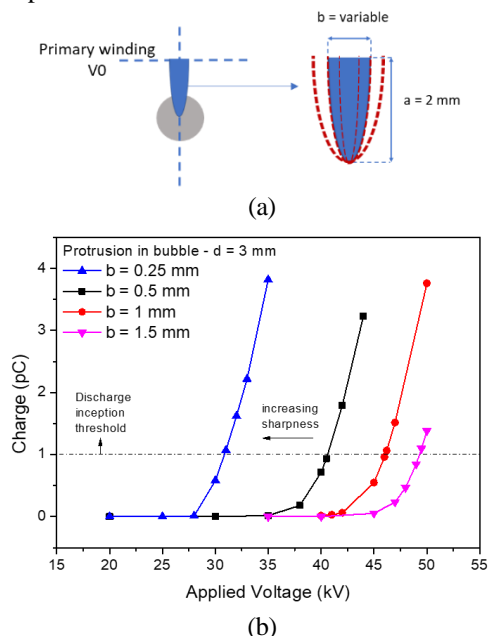

(a)

(b)

Fig. 4. (a) Computational approach for Set A - Case 3 where a gas bubble of fixed diameter 3 mm is placed inside the dielectric fluid and in contact with a protrusion in the primary winding that has variable sharpness as defined by the parameter-b of the ellipsoid geometry considered. (b) Total charge of the positive pulse as a function of the applied peak voltage $V_0$ for the different b-values of the ellipse.

Next case (Case 3) explores the effect of protrusion sharpness on the PD inception voltage when the protrusion is in direct contact with the bubble. We keep the a-value of the ellipse constant at 2 mm and the bubble diameter fixed at 3 mm. The b-value varies from 0.25 mm (very sharp tip) to 1.5 mm (close to spherical shape - typical wiring dimensions). Figure 4a shows the geometrical approach. The sharpness is critical as the electric field (EF) is maximized at the protrusion tip and the sharper the protrusion the higher the local EF and thus the higher probability for discharge inception inside the bubble. The different sharpness is translated to significant differences in the inception voltage as observed in Figure 4b. In the case of b = 1.5 mm (least sharp protrusion), the PD inception voltage was determined at 49.5 kV. This value decreases significantly when the protrusion becomes sharper, reaching values as low as 31 kV when the b-value becomes 0.25 mm. In those cases where there is a contact between the gas phase ($H_2$ plasma) and the metallic protrusion (metal contact), the discharge shifts from a rather uniform silent discharge (dielectric barrier discharge configuration) when the bubble is immersed in the dielectric fluid to a corona discharge initiated at the tip of the protrusion where the electric field becomes very high. Any sharp defects in the windings should be eliminated as the probability for PD inception even when small gas volumes come in contact becomes significantly high and at quite low transient overvoltage.

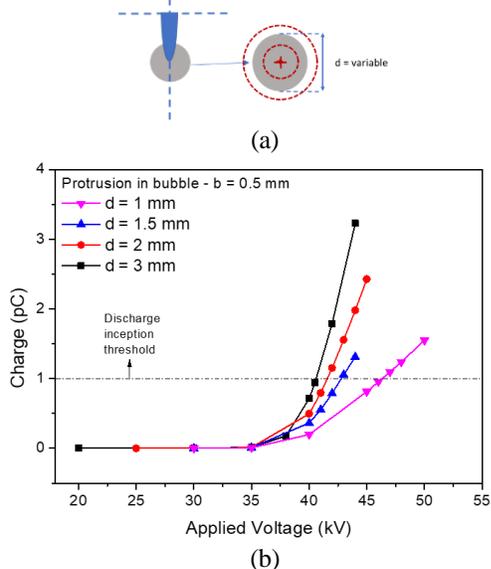

(a)

(b)

Fig. 5. (a) Computational approach of Set A – Case 4 where a gas bubble of variable diameter is placed inside the dielectric fluid and in contact with a protrusion in the primary winding that has fixed sharpness. (b) Total charge of the positive pulse as a function of the applied peak voltage $V_0$ for the different bubble diameters.

In the last case of Set A (Case 4) we also investigate the direct contact of bubble and protrusion and we explore the effect of bubble size while keeping fixed the protrusion sharpness. Namely, we fix the b-value at 0.5 mm and we change the bubble diameter from 1 to 3 mm (Figure 5a). The increase of the bubble diameter for a fixed protrusion is translated to a higher volume and a higher interface between the gas phase and the metallic electrode. Thus, a larger bubble is expected to facilitate the discharge inception. Figure 5b shows the variation of the total charge with the applied voltage for the different bubble diameters. The variations in the inception voltage are not very remarkable, however we observe a slight drop of the inception voltage value from 1 to 3 mm diameter, with the latter indicating PD inception close to 40 kV. These results are in line with Case 1 where the bubble was placed in the middle of the gap and at no contact with any metallic surface.

*B. Set B: LF/quasi-DC simulations in COPAIER*

We consider a variable-diameter bubble inside the transformer oil and in the middle of the gap between the primary and secondary winding. We do not consider the existence of any protrusion as we have found in the previous part that in the middle of the gap (far from the metallic surfaces) any protrusion in the primary winding does not affect the inception voltage. In this numerical campaign, we consider 66 kV operating RMS voltage (peak value 93 kV).

We set again the critical PD charge for PD inception to be $Q_{inc}$ =10 pC. Here, we also define and identify the *instantaneous* discharge onset (inception) as the time-instant/instantaneous applied voltage when the slope of the discharge (conduction) current changes (increases) from a smoother "pre-breakdown" like regime to a breakdown/avalanche pulsating (or not) regime and continuous growing. This roughly corresponds to maximum electron densities during the pulse peaks of $10^{13}$ - $10^{14}$ m$^{-3}$. As we are interested mainly in the instantaneous onset voltages and overall assessment of PD ignition under the 66 kV RMS applied voltage, the calculations were performed until the initiation of the PD and short-term development till the maximum positive peak voltage. We do not consider the repetition of PDs in subsequent AC half-cycles. We reiterate here that our definition of the inception voltage is a strict one, as slight deviations from the applied waveform might shift the inception voltage ($V_{inc}$) to slightly larger values and prevent total breakdown and/or partial discharging. In addition, the critical PD charge chosen here is quite strict too – a PD of around 10 pC can be rather weak and thus not detrimental for the SST operation. We consider here that the initiation of an even weak PD presents a hazard, as slight overvoltage could easily intensify the discharge which might end up into complete breakdown. A safe margin of ±1 kV around the inception voltage is suggested, considering that we do not model the superimposed HF-AC waveform as explained before.

The application of a 50 Hz voltage poses severe numerical challenges for parametrization, as the rising time of the positive-going AC phase lasts 5 ms. The extremely multi-scale nature of the problem is easily understandable if we consider that ionization time-scales range in the several ns regime while drift-diffusion timescales lie typically in the μs regime. It is thus important to find ways (retaining accuracy on inception voltage) to reduce the CPU time of the calculations to perform these highly demanding simulations. Thus, we consider that a quasi-DC case with an applied voltage ramped up to its peak value with a long enough rising time should emulate the very LF 50 Hz AC signal until the positive-going peak value. This assumption needs proper verification and thus we start by studying 4 cases: the 50 Hz AC signal and quasi-DC signals with rising times of 0.01, 0.1 and 1 ms under positive polarity. The quasi-DC cases (roughly) correspond to AC frequencies of 25 kHz, 2.5 kHz and 250 Hz respectively. For these preliminary studies, we consider as a reference case, a bubble diameter of 3 mm positioned at 25 mm from the primary winding to study the effects of the ramp-up voltage rising time in the discharge development. Figure 6 shows the instantaneous evolution of the discharge current vs the instantaneous applied voltage (current-time plot) for all cases considered.

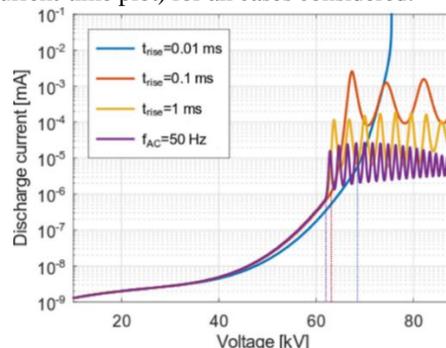

Fig. 6. Discharge current vs instantaneous applied voltage for different voltage waveforms. Dashed lines indicate the instantaneous

inception voltage. For the 50 Hz case, VRMS = 66 kV.

We note two interesting facts: 1) The instantaneous inception voltage (the applied voltage value when the current pulses start to occur) calculated for the 50 Hz case, is 62 kV. It coincides almost completely with the calculated value for the $t_{rise}$=1 ms case while it deviates by less than 1.5 kV from the $t_{rise}$=0.1 ms case. In contrast, in the $t_{rise}$=0.01 ms case the inception voltage is around 69 kV. This indicates that cases with $t_{rise}$>0.1 ms would be sufficient to accurately capture the inception voltage with a safety margin of ±1 kV, and the quasi-DC approach is justified. 2) The discharge current shows different characteristics for all cases both in its temporal behavior and its magnitude. The current for the $t_{rise}$=0.1, $t_{rise}$=1 ms and 50 Hz cases show multiple pulses which indicates that we enter a pulsating "corona-like"/PD regime. The pulsation frequency is quite different though and it increases as the applied voltage frequency increases. Moreover, the discharge current peaks are weaker for higher applied voltage frequencies. While the pulsation is not of interest in terms of the inception voltage, the current peak values (and thus the PD charge) are important (see also discussion above for the inception voltage definition) in terms of PD severity. As said, we have set the critical PD charge for PD inception to be 10 pC. The calculated PD charge for the $t_{rise}$=0.1, $t_{rise}$=1 ms and 50 Hz cases and until all waveforms reach the peak positive voltage is 16.56, 16.29 and 12 pC. Despite these values being relatively close, we choose a middle-ground between CPU efficiency and inception voltage prediction accuracy, by performing all subsequent simulations with the $t_{rise}$=1 ms case.

Lastly, it is interesting to note that for small rising times ($t_{rise}$=0.1 ms case), the discharge quickly transits to a streamer regime (indicated by the rapid increase of the current). A detailed discussion of these interesting fundamental plasma phenomena (pulsation frequency, corona-to-streamer transition) falls out of the scope of this study, although a brief discussion is provided below. We also note that in all cases, the current pulses stop when the applied voltage reaches its constant value (for the quasi-DC cases) and the current diminishes significantly to a constant lower value. For the 50 Hz AC case, the discharge diminishes as the AC waveform enters the decaying phase of the half-cycle.

Next, we perform a numerical parametric campaign for different bubble diameters: 1, 1.5, 5, 3 and 4.5 mm. For the case of 3 mm we studied both negative and positive polarity, while the bubble center position relative to the primary winding was varied (6.25, 12.5, 25 mm). For all the other cases the polarity was positive and the bubble center position fixed at 25 mm.

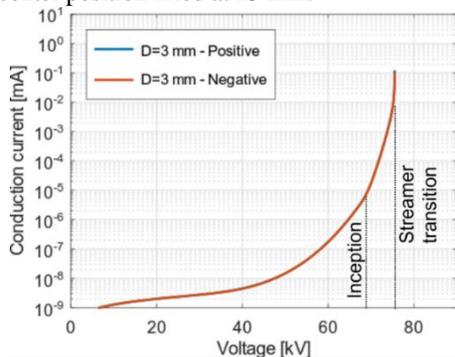

Fig. 7. Time-dependent discharge (conduction) current (A, absolute value) vs instantaneous applied voltage (kV, absolute value) for the quasi-DC case, for positive and negative polarity (curves coincide). D = 3 mm, positioned at g=25 mm. Rising time 10 μs.

Before proceeding to the parametric studies with the $t_{rise}$=1 ms case, we investigate the effect of the applied voltage polarity on the inception voltage. For this case, we still consider a bubble of 3 mm positioned at 25 mm from the primary winding and the $t_{rise}$=0.1 ms (as we are not per se interested in quantitative information here). In both positive and negative polarities, the discharge behavior is almost identical as seen in Figure 7 (both current lines coincide), and the calculated instantaneous inception voltage is $|V_{inc}|$ = 69 kV. This is expected due to the symmetry of the bubble and the domain and the absence of negative ions. In the same figure, the instantaneous inception voltage and the instantaneous voltage where the glow-like discharge transits to a streamer discharge are marked. The discharge/current behavior in repetitive AC cycles (and/or half-cycles) should be different from the first AC half-cycle: surface charging of the bubble surface should modify the potential lines and could modify both the instantaneous inception voltage (typically to lower values) and the overall discharge stability (typically stabilizing the discharge and preventing arcing). This phenomenon is typical of a DBD – the gas bubble inside a dielectric liquid closely resembles a DBD operation. The repetitive nature of PDs are important parameters for the overall safety of the device, but once a single PD is ignited at the first positive-going AC phase we can safely assume that at least one additional PD will ignite during the negative going-phase with a burst interval less than the AC half-period i.e., less than 2.5 ms. This is already close to the critical burst interval value of 2 ms.

We can now focus safely on positive polarity only. The next test case was to consider the different positioning of the gas bubble relative to the primary winding. We have performed simulations for different positions (g=25 mm, g=12.5 mm, g=6.25 mm) and the results were identical to the reference case shown above. This is expected as due to the uniformity of the electric field in the plane-to-plane configuration considered (the Laplacian electric field is relatively uniform and of same magnitude inside the gas bubble regardless of its position). This has also been validated by the results obtained in the previous section. For the remaining cases, we focus on positive polarity only and a gas bubble positioned at the middle of the domain (g=25 mm), investigating the influence of the gas bubble size (diameter). Figure 8 shows the instantaneous I-V curves for all cases considered. It is evident that the bubble dimensions play a crucial part in the inception voltage and partial discharging. Our results indicate that the larger the gas bubble, the lower the instantaneous inception voltage and the higher the total charge (and current) per PD event. Figure 9 shows the calculated charge for each case (temporal integral of the discharge current).

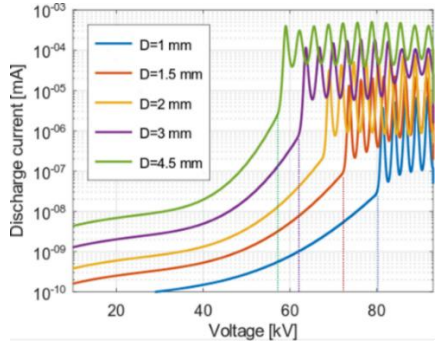

Fig. 8. Time-dependent discharge (conduction) current vs instantaneous applied voltage for the quasi-DC case and different bubble diameters. Dashed lines indicate the inception voltages.

Our results suggest that bubbles larger than 2.5 mm could initiate PDs under the operating conditions of the SST and thus it is critical to ensure that no gas impurities exist inside the dielectric oil and/or no evaporation occurs. Special attention should be paid to large bubbles, but even small bubbles can ignite a PD considering the LV-HF-50 kHz superimposed signal and overvoltage arising from external sources. It is worth noting also that larger bubbles present significantly lower instantaneous inception voltages. We can extrapolate this fact and claim that larger bubbles reduce the overall inception voltage (RMS value).

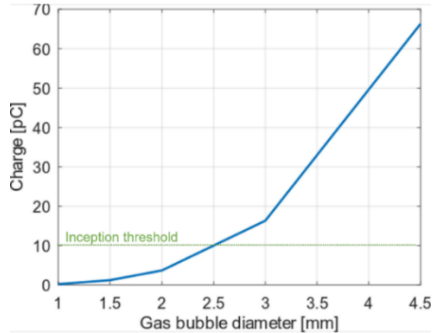

Fig. 9. Total charge per PD event versus gas bubble diameter. The inception threshold set to 10 pC is denoted with the green dashed line.

To understand the reason behind the gas bubble size influence on the inception voltage and PD charge, we have calculated the electric field enhancement factor, defined as the ratio between the maximum electric field magnitude inside the bubble and the average electric field in the dielectric fluid, before the current pulses). The calculated enhancement factor is very similar for all cases with a value of approx.1.285 (the differences are almost negligible). This denotes that the gas bubble size influence is not linked to a pure electrostatic effect – the size of the bubbles in the range considered here do not translate to a significantly different potential distribution in the domain.

We speculate that the reason behind this observation is mostly related to the characteristic charge gain/loss time scales. In small bubbles, most of the electrons (but also ions) quickly drift toward the surface of the bubble where they are lost (by surface recombination) and/or accumulated as surface charges. The loss processes are thus more dominant which inhibits charge accumulation inside the bubble and a strong avalanche mechanism.

In larger bubbles, charged species reside longer inside the bubble before drifting/diffusing away and thus the instantaneous inception voltage drops leading to current pulses of higher intensity. In Figure 11, we plot the time-averaged $H_3^+$ and electron number density for all diameters studied. The average values are quite similar but the spatial extension of the space charge is different. In small bubbles (e.g. 1mm), both high-density ion and electron regions exist only inside a portion of the bubble while for bigger bubbles (e.g. 4.5 mm) these regions are extended. Considering the bubbles' sizes, this denotes a more efficient ionization for larger bubbles inside a larger volume. This also explains the largest currents (and charge) calculated for larger bubbles.

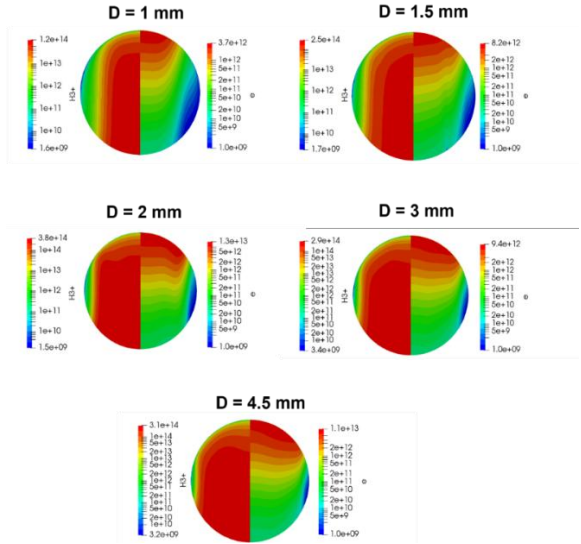

Fig. 11. Time-averaged $H_3^+$ (left half-side) and electron (right half-side) number density contours (log scale) for different bubble sizes. All figures have been scaled to the same size.

Lastly, it is worth investigating the dynamics of the PD when a corona-to-streamer transition occurs. The transition will occur for any AC frequency (or rising time value) by increasing the applied peak voltage. We thus focus on the already simulated case with a rising time of 0.01 ms (corresponding AC frequency 25 kHz), and the D=4.5 mm case which is easier to comprehend based on the numerical outputs (the rest of the cases are similar). COMSOL failed to converge in this case despite the mesh refinement used and various numerical schemes and stabilization techniques we have tried. Thus, COPAIER proved again its robustness and accuracy in such a highly demanding case. Figure 12 shows the time evolution of the electron number density during the discharge development. Due to the plane-to-plane geometry, the potential gradient inside the bubble is linear and thus the electric field is uniform. Thus, electrons are initially produced uniformly inside the bubble domain but they quickly drift towards its upper side (electrons drift in the opposite direction of the electric field vector).

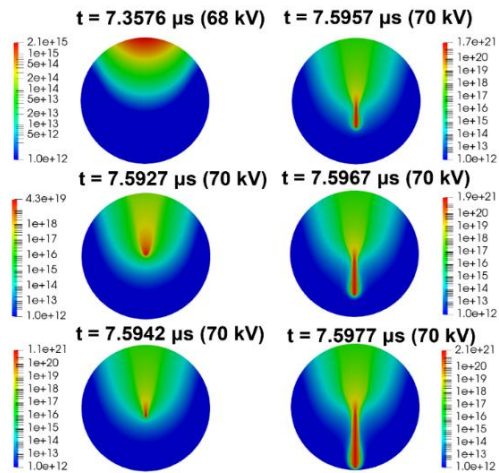

Fig. 12. Electron number density (m$^{-3}$, log-scale) contours at different time instants after PD inception (applied voltage is also noted). D = 4.5 mm, positioned at g=25 mm. The glow-like discharge inside the bubble quickly transits to a streamer discharge that bridges the gap.

At around 68 kV, electrons accumulate towards the upper side of the bubble reaching densities higher than 10$^{15}$ m$^{-3}$. The higher population of electrons at this region, enhances the electron-impact ionization reaction (R1) which leads to the production of H$_2^+$ ions (not shown here). The latter quickly transform to H$_3^+$ ions (and H atoms) via R2. Note that the ions drift velocity is several orders of magnitude lower than the electrons one. While electrons are quickly lost on the upper bubble surface (contributing to its surface charging), H$_3^+$ ions slowly drift and diffuse towards the bottom side of the bubble leaving behind a positive space charge which grows in time and disturbs the external electric field. At ~70 kV (t=7.5927 µs), this space charge located near the bubble center along with the conductive nature of the (rather weak still) quasi-neutral plasma formed near the top side of the bubble leads to a strong electric field enhancement, which causes the corona-to-streamer transition. The streamer propagates in the form of an ionization wave and in less than 5 ns it bridges the bubble gap. Its propagation speed is estimated to be ~0.5 mm/ns. The streamer body holds electron densities higher than 10$^{21}$ m$^{-3}$. A thin "cathode layer" is formed upon impingement to the bottom side of the bubble. In later times, the streamer is expected to slowly relax and/or transit to a surface ionization wave propagating along the bubble surface and guided by surface charging effects. At the time of impingement, the dominant positive species are H$_3^+$ ions (not shown here for brevity).

## V. CONCLUSION

We used self-consistent plasma modeling to investigate the PD occurrence at the most critical parts of the SST module which is subjected to extremely high potential differences. We have explored different scenarios of PD inception considering a H$_2$ bubble trapped inside transformer oil. We also studied the effect of additional defects, i.e., a metallic protrusion of a high local electric field. The numerical results showed that gas bubbles can be detrimental for the insulating system, however for transient overvoltage the occurrence of PD is less probable for bubbles floating in the dielectric fluid if there is no defect (protrusion) on the winding surface. The smaller the bubble the less the inception voltage, and for mm-size bubbles when 50 kHz signals are applied the peak voltages for inception are >70 kV. The existence of metallic protrusion can affect the inception voltage of a remote floating bubble only slightly and when this is close to the sharp tip, with inception voltage remaining still >70 kV. The extreme scenario of a protrusion in contact with a bubble severely affects the insulation properties and drops the PD inception voltage remarkably. The larger the bubble and the sharper the tip of the protrusion the lower the inception peak voltage, which can reach values well below 40 kV. It is thus critical to keep the winding surfaces and wiring as smooth as possible, while gas impurities should be minimized. The numerical results also suggest that gas bubbles pose a significant hazard for PD event occurrence even at LF high voltages. Larger bubbles increase the severity and probability of PD events. They also lead to lower instantaneous inception voltages. Current pulses produced in bubbles can quickly transit to intense streamer discharges (which can also transit to catastrophic arcing) if the frequency is reduced and/or under transient, HF overvoltage.

The simulations provide basic guidelines on how defects trigger PD probabilities in high-voltage transformers and lay the ground for more detailed computational studies in the field. We note that possible deformation of bubbles and/or their movement (bubbles are considered immobile) due to external fluid flow and/or electric (Coulomb) forces could play an important role in PD inception. In the future, the proposed modeling procedure could be coupled in a multi-physics simulation module to couple the plasma dynamics with bubble deformation, fluid and thermo-dynamics and investigate the complex interactions arising from the multiple transport and electrostatic phenomena inside the transformer devices.

## ACKNOWLEDGMENTS

This work was funded in the framework of SSTAR (Innovative HV Solid-State Transformer for maximizing renewable energy penetration in energy distribution and transmission systems) Horizon Europe project (Grant agreement ID: 101069702).

## REFERENCES

[1] W. J. K. Raymond, H. A. Illias, A. H. A. Bakar, and H. Mokhlis, "Partial discharge classifications: Review of recent progress," *Measurement: Journal of the International Measurement Confederation*, vol. 68. Elsevier B.V., pp. 164–181, 2015, doi: 10.1016/j.measurement.2015.02.032.

[2] M. R. Hussain, S. S. Refaat, and H. Abu-Rub, "Overview and Partial Discharge Analysis of Power Transformers: A Literature Review," *IEEE Access*, vol. 9. Institute of Electrical and Electronics Engineers Inc., pp. 64587–64605, 2021, doi: 10.1109/ACCESS.2021.3075288.

[3] N. J. Felici, "Bubbles, partial discharges and liquid breakdown," *Electrost. Inst. Phys. Conf. Ser.*, vol. 79, pp. 181–190, 1979.

[4] M. S. Naidu and V. Kamaraju, *High Voltage Engineering*. Tata McGraw-Hill Publishing Company Limited, New Delhi, 2000.

[5] M. G. Danikas, "Bubbles in Insulating Liquids: A Short Review," *Eng. Technol. Appl. Sci. Res.*, vol. 9, no. 6, pp. 4870–4875, 2019.

[6] T. Buchacz, J. Buchacz, and M. Duval, "Stray Gassing of Oil in HV Transformers," *IEEE Trans. Dielectr. Electr. Insul.*, vol. 28, no. 5, pp. 1729–1734, 2021, doi: 10.1109/TDEI.2021.009520.

[7] A. J. Amalanathan, R. Sarathi, M. Zdanowski, R. Vinu, and Z. Nadolny, "Review on Gassing Tendency of Different Insulating Fluids towards Transformer Applications," *Energies*, vol. 16, no. 1, pp. 1–15, 2023, doi: 10.3390/en16010488.

[8] C. Pan, G. Chen, J. Tang, and K. Wu, "Numerical modeling of partial discharges in a solid dielectric-bounded cavity: A review," *IEEE Trans. Dielectr. Electr. Insul.*, vol. 26, no. 3, pp. 981–1000, Jun. 2019, doi: 10.1109/tdei.2019.007945.

[9] "S. Whitehead, Dielectric breakdown of solids, Clearendon Press, 1952."

[10] A. Pedersen, G. C. Crichton, and I. W. McAllister, "The Functional Relation between Partial Discharges and Induced Charge," *IEEE Trans. Dielectr. Electr. Insul.*, vol. 2, no. 4, pp. 535–543, 1995, doi: 10.1109/94.407019.

[11] Z. Achillides, G. E. Georghiou, and E. Kyriakides, "Partial Discharges and Associated Transients: The Induced Charge Concept versus Capacitive Modeling," 2008.

[12] H. A. Illias, G. Chen, and P. L. Lewin, "The influence of spherical cavity surface charge distribution on the sequence of partial discharge events," *J. Phys. D. Appl. Phys.*, vol. 44, no. 24, Jun. 2011, doi: 10.1088/0022-3727/44/24/245202.

[13] P. D. P. B. G. S. G. V. B.-C. M.M. Gongora-Nieto, "Impact of air bubbles in a dielectric liquid when subjected to high field strengths," *Innov. Food Sci. Emerg. Technol. 4 57–67*.

[14] N. Y. Babaeva, D. V. Tereshonok, and G. V. Naidis, "Initiation of breakdown in bubbles immersed in liquids: Pre-existed charges versus bubble size," *J. Phys. D. Appl. Phys.*, vol. 48, no. 35, Sep. 2015, doi: 10.1088/0022-3727/48/35/355201.

[15] N. Pillai, N. L. Sponsel, J. T. Mast, M. J. Kushner, I. A. Bolotnov, and K. Stapelmann, "Plasma breakdown in bubbles passing between two pin electrodes," *J. Phys. D. Appl. Phys.*, vol. 55, no. 47, Nov. 2022, doi: 10.1088/1361-6463/ac9538.

[16] "Plasma Module User's Guide, www.comsol.com."

[17] K. Kourtzanidis, G. Dufour, and F. Rogier, "Self-consistent modeling of a surface AC dielectric barrier discharge actuator: In-depth analysis of positive and negative phases," *J. Phys. D. Appl. Phys.*, vol. 54, no. 4, Jan. 2021, doi: 10.1088/1361-6463/abbcfd.

[18] C. Besse, T. D. Nguyen, F. Rogier, and N. T. Dung, "An implicit time integration approach for simulation of corona discharges." [Online]. Available: https://hal.science/hal-04102798.

[19] G. J. M. Hagelaar and L. C. Pitchford, "Solving the Boltzmann equation to obtain electron transport coefficients and rate coefficients for fluid models," *Plasma Sources Sci. Technol.*, vol. 14, no. 4, pp. 722–733, Nov. 2005, doi: 10.1088/0963-0252/14/4/011.

[20] J. S. Yoon *et al.*, "Cross sections for electron collisions with hydrogen molecules," *J. Phys. Chem. Ref. Data*, vol. 37, no. 2, pp. 913–931, 2008, doi: 10.1063/1.2838023.

[21] Y. R. Zhang, X. Xu, A. Bogaerts, and Y. N. Wang, "Fluid simulation of the phase-shift effect in hydrogen capacitively coupled plasmas: I. Transient behaviour of electrodynamics and power deposition," *J. Phys. D. Appl. Phys.*, vol. 45, no. 1, Jan. 2012, doi: 10.1088/0022-3727/45/1/015202.

[22] A. Salabaş and R. P. Brinkmann, "Numerical investigation of dual frequency capacitively coupled hydrogen plasmas," *Plasma Sources Sci. Technol.*, vol. 14, no. 2, May 2005, doi: 10.1088/0963-0252/14/2/S07.

[23] T. Farouk, B. Farouk, D. Staack, A. Gutsol, and A. Fridman, "Modeling of direct current micro-plasma discharges in atmospheric pressure hydrogen," *Plasma Sources Sci. Technol.*, vol. 16, no. 3, pp. 619–634, Aug. 2007, doi: 10.1088/0963-0252/16/3/023.

[24] B. Kalache, T. Novikova, A. Fontcuberta i Morral, P. Roca i Cabarrocas, W. Morscheidt, and K. Hassouni, "Investigation of coupling between chemistry and discharge dynamics in radio frequency hydrogen plasmas in the Torr regime," *J. Phys. D. Appl. Phys.*, vol. 37, no. 13, pp. 1765–1773, Jul. 2004, doi: 10.1088/0022-3727/37/13/007.

[25] "Morgan database, www.lxcat.net, retrieved on October 5, 2022."

[26] E. Carbone *et al.*, "Data needs for modeling low-temperature non-equilibrium plasmas: The LXCat project, history, perspectives and a tutorial," *Atoms*, vol. 9, no. 1. MDPI AG, pp. 1–40, Mar. 01, 2021, doi: 10.3390/atoms9010016.

[27] E. A. Mason and J. T. Vanderslice, "Mobility of Hydrogen Ions ($H^+$, $H_2^+$, $H_3^+$) in Hydrogen*."

[28] C. Geuzaine and J.-F. Remacle, "Gmsh: a three-dimensional finite element mesh generator with built-in pre-and post-processing facilities," 2009.

[29] U. Schichler *et al.*, "Risk assessment on defects in GIS based on PD diagnostics," *IEEE Trans. Dielectr. Electr. Insul.*, vol. 20, no. 6, pp. 2165–2172, 2013, doi: 10.1109/TDEI.2013.6678866.

[30] M. Alshaikh Saleh, S. S. Refaat, M. Olesz, H. Abu-Rub, and J. Guziński, "The effect of protrusions on the initiation of partial discharges in XLPE high voltage cables," *Bull. Polish Acad. Sci. Tech. Sci.*, vol. 69, no. 1, Feb. 2021, doi: 10.24425/bpasts.2021.136037.

[31] Y. Peng, C. Feng, Y. Song, and C. Min, "Single Bubble Behavior in Direct Current Electric Field *," 2008.

[32] M. Ghaffarian Niasar, H. Edin, X. Wang, and R. Clemence, "PARTIAL DISCHARGE CHARACTERISTICS DUE TO AIR AND WATER VAPOR BUBBLES IN OIL."

[33] R. Peñaloza-Delgado, J. L. Olvera-Cervantes, M. E. Sosa-Morales, T. K. Kataria, and A. Corona-Chávez, "Dielectric characterization of vegetable oils during a heating cycle," *J. Food Sci. Technol.*, vol. 58, no. 4, pp. 1480–1487, Apr. 2021, doi: 10.1007/s13197-020-04660-7.